\documentclass[12pt]{iopart}
\usepackage{graphicx}
\usepackage{dcolumn}
\usepackage{bm}
\usepackage[tight]{subfigure}
\usepackage{verbatim}
\usepackage{color}

\begin{document}
\title{Thermoelectric unipolar spin battery in a suspended carbon nanotube}
\author{Zhan Cao$^1$, Tie-Feng Fang$^1$, Wan-Xiu He$^1$, and Hong-Gang Luo$^{1,2}$}
\address{$^1$ Center for Interdisciplinary Studies $\&$ Key Laboratory for Magnetism and Magnetic Materials of the Ministry of Education, Lanzhou University, Lanzhou 730000, China}
\address{$^2$ Beijing Computational Science Research Center, Beijing 100084, China}

\begin{abstract}
A quantum dot formed in a suspended carbon nanotube exposed to an external magnetic field is predicted to act as a thermoelectric unipolar spin battery which generates pure spin current. The built-in spin flip mechanism is a consequence of the spin-vibration interaction resulting from the interplay between the intrinsic spin-orbit coupling and the vibrational modes of the suspended carbon nanotube. On the other hand, utilizing thermoelectric effect, the temperature difference between the electron and the thermal bath to which the vibrational modes are coupled provides the driving force. We find that both magnitude and direction of the generated pure spin current are dependent on the strength of spin-vibration interaction, the sublevel configuration in dot, the temperatures of electron and thermal bath, and the tunneling rate between the dot and the pole. Moreover, in the linear response regime, the kinetic coefficient is non-monotonic in the temperature $T$ and it reaches its maximum when $k_BT$ is about one phonon energy. The existence of a strong intradot Coulomb interaction is irrelevant for our spin battery, provided that high-order cotunneling processes are suppressed.
\end{abstract}

\maketitle
\section{Introduction} \label{intr}
Generating spin current is one of the fundamental issues in spintronics \cite{Wolf2001,Zutic2004}. When spin-up and spin-down electrons travel in opposite directions, the net charge current is $I_c\equiv e(J_{\uparrow}+J_{\downarrow})$ while the spin current is $I_s\equiv\hbar/2(J_{\uparrow}-J_{\downarrow})$, where $J_\uparrow$ ($J_\downarrow$) is the spin-up (spin-down) electron current. A device which can drive a spin current into external circuits is called spin battery (SB) \cite{Sun2003,Long2003,Brataas2002,Wang2004}. Thus far, various SBs have been proposed, e.g., the earlier dipolar and unipolar SBs summarized in reference \cite{Wang2004}, and the following three-terminal devices consisting of metallic/ferromagnetic poles \cite{Pareek2004,Chen2004,Wang2005,Lu2007,Nazarov2007} or even involving superconducting pole \cite{Futterer2010,Wysokinski2012}. Among the existing schemes, in most of the multipolar SBs, both charge current and spin current coexist except for particular parameter regimes. However, the charge current in a unipolar SB must be zero in the steady state since only one pole exists \cite{Wang2004}. From this perspective, the unipolar SB is the superior candidate for generating a pure spin current (PSC) with $I_c=0$ and $I_s\ne0$. A built-in spin flip mechanism is usually necessary in a unipolar SB, namely, it draws in electrons with one spin orientation from the pole, then flips the spin inside the SB, followed by pushing out the electron with opposite spin orientation to the pole \cite{Wang2004}. If this kind of SB is connected to an external circuit which is not closed, it drives a PSC in that circuit. Typical built-in spin flip mechanism in a unipolar SB is a ferromagnetic resonance or a rotating external magnetic field \cite{Brataas2002,Zhang2003,Wang2003}, which involves somewhat complicated time-varying external fields.

\begin{figure}[bp]
\includegraphics[width=0.9\columnwidth]{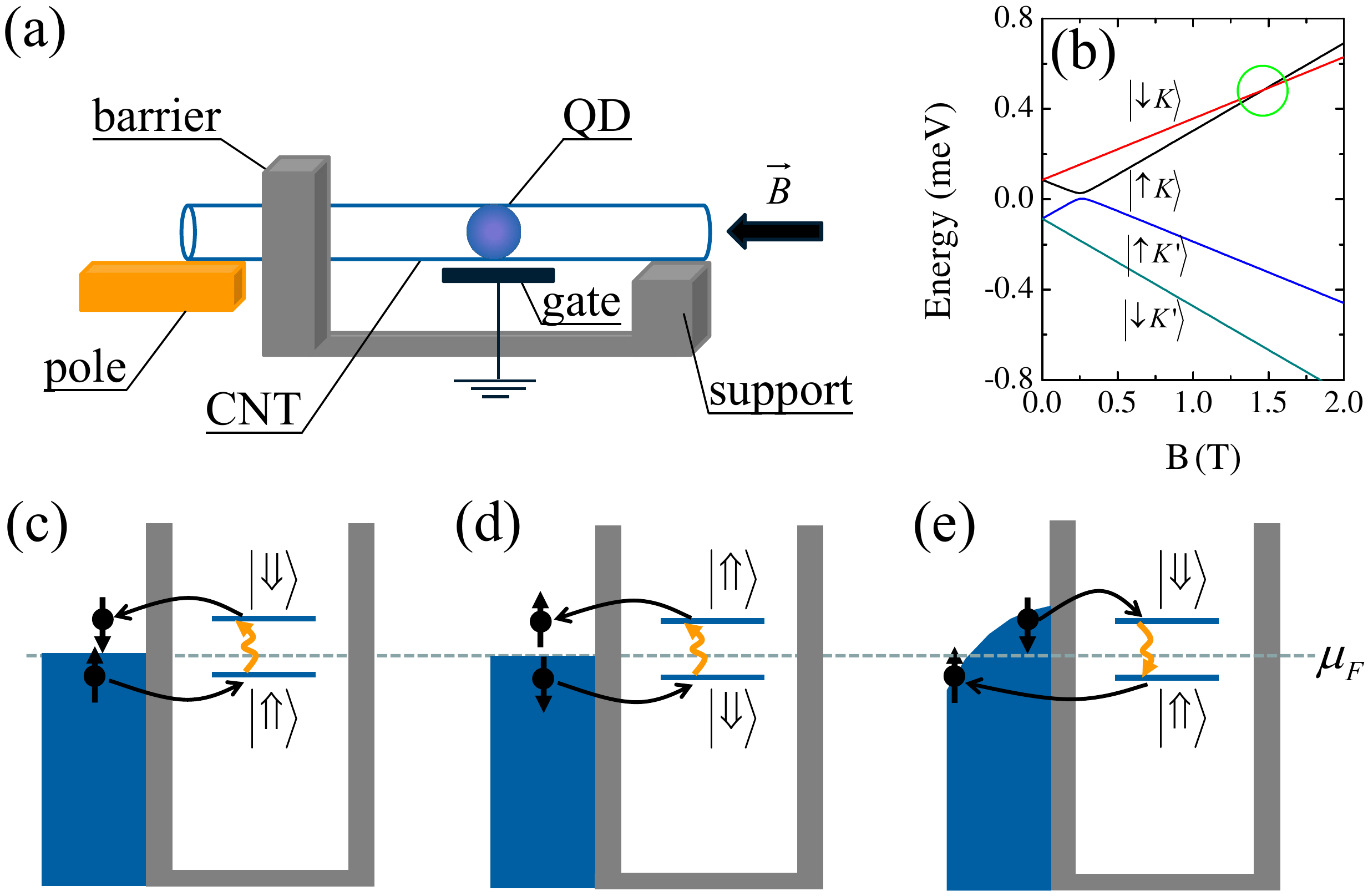}
\centering
\caption{(Color online) (a) Schematic of a SB setup made up of a suspended CNT containing a QD with split energy level due to applied longitudinal magnetic field $\vec B$. (b) Eigenenergies of a CNT QD as a function of the magnitude of $\vec B$ with $E_0=0$ and the CNT parameters $\Delta_{so}=170\mu eV$, $\Delta_{KK'}=12.5\mu eV$, and $\mu_{orb}=330\mu eV/T$. Here, the values of $\Delta_{so}$, $\Delta_{KK'}$, and $\mu_{orb}$ are extracted from reference \cite{Churchill2009}. The circle marks the exact crossing point $E_{\uparrow K}=E_{\downarrow K}$. (c)-(e) illustrate the mechanism of generating PSCs. Electron temperature $T_e$ in (c) and (d) is zero while in (e) is finite.}\label{fig1}
\end{figure}

In this work, we predict a unipolar SB made up of a quantum dot (QD) formed in a suspended carbon nanotube (CNT) exposed to an external magnetic field, see figure \ref{fig1}(a). As we shall see below, instead of employing a time-varying external field, a natural spin flip source is available in this SB setup. It is known that in a CNT QD, the twofold spin and twofold orbital symmetries give rise to a fourfold degenerate energy level $E_0$. In the presence of the spin-orbit coupling $\Delta_{so}$, intervalley scattering $\Delta_{KK'}$, and applying an external longitudinal magnetic field $\vec B$ which couples to the electronic orbital ($\mu_{orb}$) and spin ($\mu_B$) magnetic moments, the degenerate energy level $E_0$ splits into four branches \cite{Rudner2010} $E_{\uparrow(\downarrow) K}=E_0\pm\mu_B B+\frac {1}{2}\sqrt{(\Delta_{so}\mp2\mu_{orb}B)^2+4\Delta_{KK'}^2}$ and $E_{\uparrow(\downarrow) K'}=E_0\pm\mu_B B-\frac {1}{2}\sqrt{(\Delta_{so}\mp2\mu_{orb}B)^2+4\Delta_{KK'}^2}$, as shown in figure \ref{fig1}(b). Experimentally, by the cooperation of a gate voltage (controlling $E_0$) and a longitudinal magnetic field, the four-level structure can be finely tuned \cite{Jarillo2005,Kuemmeth2008,Fang2008,Churchill2009}. In particular, close to the exact crossing point $E_{\uparrow K}=E_{\downarrow K}$ (marked by the circle in figure \ref{fig1}(b)), one has two levels of opposite spin and the same orbital and their energy separation is smaller than the energy distance from other levels. Recently, a phonon-mediated spin flip mechanism established within this two-level subspace has been addressed \cite{Ohm2012,Palyi2012,Stadler2014,Stadler2015}. It is a consequence of the spin-vibration interaction (SVI) resulting from the interplay between the intrinsic spin-orbit coupling and the vibrational modes of the suspended CNT. The coupling constant of SVI reads $\lambda_n\simeq(\Delta_{so}/2)\mu_n\langle df_n(z)/dz \rangle$ \cite{Palyi2012,Stadler2014}, with $f_n(z)$ the profile function, and $\mu_n=\sqrt{\hbar/(2m\omega_n)}$ with $m$ and $\omega_n$ being the nanotube's mass and the $n$th eigenfrequency of the vibration, respectively. One can estimate $\lambda \sim 2.5$MHz for the first odd mode in a typical CNT \cite{Stadler2014}. Recent experiment suggests an even stronger $\lambda$ due to the large $\Delta_{so}$ measured \cite{Steele2013}. Moreover, the vibrational modes of CNTs usually couple to an equilibrium thermal phonon bath representing the environment induced by thermal nuclear motions \cite{Galperin2007,Zimbovskaya2011}. It has been found that, at unequal electron ($T_e$) and thermal bath ($T_b$) temperatures, the heat current between the bath and the electrons can be converted into an electron current \cite{Entin2010}. We appreciate that this nontrivial thermoelectric effect could provide the driving force needed in the present SB setup. It is worth mentioning that the nonequilibrium between $T_e$ and $T_b$ in CNTs has already been observed experimentally \cite{Lazzeri2005,Carl2008,Berciaud2010} and discussed theoretically \cite{Zippilli2009,Dubi2011,Fang2011}. In particular, the bath temperature is maintained provided that the SVI strength $\lambda$ is much weaker (which is true in a typical CNT \cite{Palyi2012}) than the coupling of CNT vibrations to the thermal phonon bath \cite{Entin2010}. Alternatively, an artificial thermal phonon bath held at a temperature $T_b$ may be realized simply by an electronically insulating hard substrate touching the quantum dot \cite{Entin2010}, which is spatially well separated from the electronic pole. Hence, a stable temperature difference can be achieved by heating or cooling the electronic pole solely.

First of all, we illustrate why the setup sketched in figure \ref{fig1}(a) can work as a SB. For a large CNT QD without Coulomb interaction, the physics is summarized in figures \ref{fig1}(c)-\ref{fig1}(e). As shown in figure \ref{fig1}(c), at $T_e=0$ the Fermi distribution in the external pole changes abruptly from unity to zero at the Fermi level $\mu_F$. By adjusting the gate voltage together with the applied magnetic field, the QD energy level can split into two sublevels such that $\varepsilon_{d\uparrow}<\mu_F<\varepsilon_{d\downarrow}$. In this case, the lower sublevel is occupied by a spin-up electron while the upper one is empty. For nonzero $T_{b}$ where thermal phonons are available, due to the SVI, the spin-up electron at lower sublevel can reverse its spin and transit into an excited state with energy $\varepsilon^{ex}_{d\downarrow}=\varepsilon_{d\uparrow}+\varepsilon_{ph}$ by absorbing a phonon. Provided that $\varepsilon^{ex}_{d\downarrow}>\mu_F$, this excited spin-down QD electron can easily tunnel out to the external pole. During this combined process, the spin-up electrons continuously flow out of the pole into the QD while the spin-down QD electrons persist in injecting to the pole, which successfully establishes a positive PSC ($J_\downarrow<0<J_\uparrow$). When the resonance condition $\varepsilon_{d\downarrow}-\varepsilon_{d\uparrow}=\varepsilon_{ph}$ (i.e., $\varepsilon^{ex}_{d\downarrow}=\varepsilon_{d\downarrow}$) is satisfied, the magnitude of PSC reaches its maximum. Similarly, a negative PSC ($J_\uparrow<0<J_\downarrow$) can be achieved under the opposite sublevel configuration $\varepsilon_{d\downarrow}<\mu_F<\varepsilon_{d\uparrow}$ [see figure \ref{fig1}(d)]. In figure \ref{fig1}(e), when finite $T_e$ is considered the Fermi distribution is smeared around $\mu_F$ and thereby a few holes (electrons) are available below (above) $\mu_F$. As a result, the spin-up electron occupied at the lower sublevel can tunnel to the pole, while a spin-down pole electron can inject to the upper sublevel in QD and then transits immediately to the lower sublevel by emitting a phonon. This is an opposite process to the one depicted in figure \ref{fig1}(c). With this respect, the positive PSC will be suppressed at finite $T_e$. Nevertheless, this also provides a mechanism to reverse the PSC when the process in figure \ref{fig1}(e) prevails over the one in figure \ref{fig1}(c).

The inclusion of a strong Coulomb interaction in the CNT QD will not disturb the substantial physical scenario, provided that only the Coulomb blockade effect \cite{Meir1991} survives whereas all the high-order cotunneling processes \cite{Franceschi2001,Hewson1993} are suppressed at weak tunnel coupling or at high enough temperatures. Physically, a finite Coulomb repulsion $U$ will induce two more sublevels with higher energies $\varepsilon_{d\uparrow}+U$ and $\varepsilon_{d\downarrow}+U$. Nevertheless, one can always adjust the upper or lower two sublvels to the vicinity of $\mu_F$ to act as the two relevant sublevels depicted in the illustrations, while the other two sublevels deviating largely from the Fermi level do not take part in the PSC generating. This is demonstrated in section \ref{interacting} by the master equation calculations incorporating the electron-electron correlations at the Coulomb blockade level.

The scenarios proposed above indicate that the magnitude and the direction of PSC are dependent on the strength of SVI, the sublevel configuration in QD, and the electron temperature $T_e$, (or relatively, the bath temperature $T_b$). Moreover, as we discuss below, the tunneling rate between the QD and the pole, as well as the intradot Coulomb interaction, can also influence the magnitude of PSC. In what follows, we identify the physics mentioned above by solving the model Hamiltonian presented in section \ref{model}. In section \ref{noninteracting}, the nonequilibrium Green's function theory is employed to study the PSC generated in a QD without Coulomb interaction. In section \ref{interacting}, we use the master equation method to study the effect of Coulomb interaction on the PSC generation. Finally, a conclusion is given in section \ref{summary}.

\section{Model Hamiltonian}\label{model}
The SB setup we consider [figure \ref{fig1}(a)] can be described by the Hamiltonian with a general form $H=H_{pole}+H_{ph}+H_{QD}+H_{tunnel}$, where
\begin{eqnarray}
&&H_{pole}=\sum_{k,\sigma} \varepsilon_{k}c_{k\sigma}^\dag c_{k\sigma},\\
&&H_{ph}=\varepsilon_{ph} a^\dag a,\\
&&H_{QD}=\sum_{\sigma}\varepsilon_{d\sigma}d_{\sigma}^{\dag}d_{\sigma}+Un_{d\uparrow}n_{d\downarrow}+\lambda(a+a^\dag)\sum_{\sigma}d_\sigma^\dag d_{\bar\sigma},\\
&&H_{tunnel}=\sum_{k,\sigma}(t d_{\sigma}^{\dag}c_{k\sigma}+\textrm{H.c.}).
\end{eqnarray}
Here, $\varepsilon_k$ is the single-particle energy of an electron with momentum $k$ in the noninteracting external pole. $H_{ph}$ represents a single vibrational mode of frequency $\varepsilon_{ph}/\hbar$, which describes the vibration of the suspended CNT. $H_{QD}$ describes the effective two-level CNT QD influenced by the SVI \cite{Palyi2012,Stadler2014,Stadler2015}, as mentioned above. $\varepsilon_{d\sigma}$ denotes the spin-dependent sublevels, $U$ is the Coulomb repulsion energy in the QD, and $\lambda$ measures the strength of SVI. $H_{tunnel}$ stands for the tunneling coupling between the QD and the pole, with $t$ being the tunneling matrix element. An electron and/or hole transferring between the QD and the pole is described by an effective tunneling rate $\Gamma=2\pi\rho|t|^2$, where $\rho$ is the pole density of states.

\section{QD without Coulomb interaction}\label{noninteracting}
\subsection{Nonequilibrium Green's function formalism}
We first consider a large QD in which the intradot Coulomb interaction could be neglected. Using the standard Keldysh nonequilibrium Green's function theory \cite{Haug2008}, the steady spin-dependent electron current, $J_\sigma=\frac{i}{\hbar}\langle[N_\sigma,H]\rangle$ with $N_\sigma=\sum_{k}c_{k\sigma}^\dag c_{k\sigma}$, flowing through pole into the QD can be expressed as
\begin{equation}
J_\sigma=\frac{1}{\hbar}\int \frac{d\omega}{2\pi} [\Sigma_0^<(\omega)G^>_\sigma(\omega)-\Sigma_0^>(\omega)G^<_\sigma(\omega)],\label{eq1}
\end{equation}
where $G_{\sigma}^{<}(\omega)$ and $G_{\sigma}^{>}(\omega)$ represent the full lesser and greater Green's functions (GFs) of the localized QD electron. $\Sigma^{<}_0(\omega)=i\Gamma f(\omega)$ and $\Sigma^{>}_0(\omega)=i\Gamma [f(\omega)-1]$ are the lesser and greater self-energies, respectively, contributed from tunnel coupling to the pole. $f(\omega)=\{\textrm{exp}[\omega/k_B T_e]+1\}^{-1}$ is the Fermi distribution of pole electron (we set $\mu_F=0$). Here the spin accumulation (spin-dependent $\mu_F$) in the pole is neglected assuming that the size of the pole is sufficiently large and the spin-relaxation time is sufficiently short \cite{Swirkowicz2009}. It means that the generated electron current injects to the external circuit promptly.

To solve the relevant lesser and greater QD GFs, one has to make some approximations since the study on phonon-mediated inelastic transports is far from trivial, even though the QD itself is noninteracting. Various treatments on dealing with electron-phonon interaction from weak to strong coupling regime as well as from equilibrium to nonequilibrium have been established \cite{Galperin2007,Zimbovskaya2011} thus far. In this work, we focus on the weak coupling regime where $\lambda\ll\Gamma, \varepsilon_{ph}$, which is true in a typical suspended CNT device \cite{Palyi2012}. In this case, a generalization of the self-consistent Born approximation \cite{Viljas2005,Frederiksen2007}, devised for treating the Holstein-type electron-phonon interaction, is straightforward. The main idea lies in that assuming the interactions of the QD with pole and phonons are independent of each other, i.e., the total self-energies are obtained in the form $\Sigma_\sigma^{x}=\Sigma_0^{x}+\Sigma_{\sigma;ph}^{x}$, where $x=(r,a,<,>)$ with $r(a)$ denotes the retarded (advanced) component. Therefore, the lesser (greater) QD GF can be obtained by following Keldysh equation \cite{Haug2008}
\begin{equation}
G_\sigma^{<(>)}(\omega)=G_\sigma^{r}(\omega)[\Sigma_0^{<(>)}(\omega)+\Sigma_{\sigma;ph}^{<(>)}(\omega)]G_\sigma^{a}(\omega), \label{Keldysh}
\end{equation}
where
\begin{equation}
\Sigma_{\sigma;ph}^{<(>)}(\omega)=\lambda^2[N_{ph}G_{\bar\sigma}^{<(>)}(\omega\mp\varepsilon_{ph})+(1+N_{ph})G_{\bar\sigma}^{<(>)}(\omega\pm\varepsilon_{ph})] \label{se}
\end{equation}
is the lesser (greater) self-energy obtained by considering the Hartree and Fock self-energy diagrams \cite{Mahan2000}, regarding the SVI as the perturbation term. Here, $\bar\sigma$ represents the opposite spin of $\sigma$. $N_{ph}= \{\textrm{exp}[\varepsilon_{ph}/k_B T_{b}]-1\}^{-1}$ is the average number of phonons in the equilibrium thermal bath to which the vibrational mode is coupled. We note that the Hartree self-energy vanishes since there is no spontaneous spin-flip term in the present Hamiltonian $H$.

Substituting equations (\ref{Keldysh}) and (\ref{se}) into the current formula equation (\ref{eq1}), one immediately obtains
\begin{eqnarray}
J_\sigma=&&\lambda^2\frac{1}{\hbar}\int \frac{d\omega}{2\pi} G^r_\sigma(\omega)\{\Sigma_0^<(\omega)[N_{ph}G_{\bar\sigma}^{>}(\omega+\varepsilon_{ph})
+(1+N_{ph})G_{\bar\sigma}^{>}(\omega-\varepsilon_{ph})]\nonumber\\
&&-\Sigma_0^>(\omega)[N_{ph}G_{\bar\sigma}^{<}(\omega-\varepsilon_{ph})+(1+N_{ph})G_{\bar\sigma}^{<}(\omega+\varepsilon_{ph})]\}G^a_\sigma(\omega),\label{c1}
\end{eqnarray}
Within the framework of self-consistent Born approximation, all the full GFs and self-energies have to be solved in an iterative manner, however, for weak SVI as we consider, to the lowest-order of the coupling strength $\lambda$, one can replace the full GFs appear in equation (\ref{c1}) by their bare counterparts $G^{0,r(a)}_\sigma(\omega)=(\omega-\varepsilon_{d\sigma}\pm i\Gamma/2)^{-1}$ and $G_\sigma^{0,<(>)}(\omega)=G_\sigma^{0,r}(\omega)\Sigma_0^{<(>)}(\omega)G_\sigma^{0,a}(\omega)$. Collecting these terms we arrive at a compact current formula
\begin{equation}
J_{\sigma}=\lambda^{2}\frac{2\pi}{\hbar}\int d\omega\mathcal T_\sigma(\omega) \mathcal F(\omega), \label{eq6}
\end{equation}
where $\mathcal T_\sigma(\omega)=\rho_{\sigma}(  \omega_{-})\rho_{\bar{\sigma}}(  \omega_{+})-\rho_{\sigma}(  \omega_{+}) \rho_{\bar{\sigma}}(  \omega_{-})$ and $\mathcal F(\omega)  = (  1+N_{ph}) [f(\omega_{-})  -1]  f(  \omega_{+})-N_{ph}f(  \omega_{-})[  f(  \omega_{+})-1]$, with $\omega_{\pm}=\omega\pm\varepsilon_{ph}/2$. Here, $\rho_\sigma(\omega)=-\frac{1}{\pi}\textrm{Im}G^{0,r}_\sigma(\omega)$ is the spin-resolved density of dot states in the absence of SVI. The charge current conservation law $I_c\equiv e(J_\uparrow+J_\downarrow)=0$ is readily checked by replacing $\sigma$ by $\bar\sigma$ in equation (\ref{eq6}). Thus the generated spin current is indeed a PSC with $I_s=\hbar J_\uparrow$. It is not difficult to understand that this driven PSC is a consequence of the nontrivial thermoelectric effect with respect to the Fermi (pole) and Bose (thermal bath) reservoirs. Some analytically insights about equation (\ref{eq6}) are summarized as follows:

i) The magnitude of $I_s$ is proportional to $\lambda^2$.

ii) $I_s$ reverses its direction but keeps the magnitude when we exchange the positions of spin-up and spin-down sublevels.

iii) The function $\mathcal{F}(\omega)$ has exactly the same sign with $\Delta T=T_{b}-T_{e}$, irrespective of the specific parameters, and it becomes zero when $\Delta T=0$. Particularly, $\mathcal F(\omega)$ converge to a constant $-1/4$ in the limit $T_{e}\gg T_{b}$.

On the other hand, in the limit $T_{e}\rightarrow 0$, the Fermi distribution reduces to the Heaviside function such that the PSC becomes
\begin{eqnarray}
I_s=2\pi\lambda^{2} N_{ph}\int^{\varepsilon_{ph}/2}_{-\varepsilon_{ph}/2} d\omega \mathcal T_\uparrow(\omega). \label{eq7}
\end{eqnarray}
\subsection{Numerical results and discussions}\label{results}
\begin{figure}[bp]
\includegraphics[width=\columnwidth]{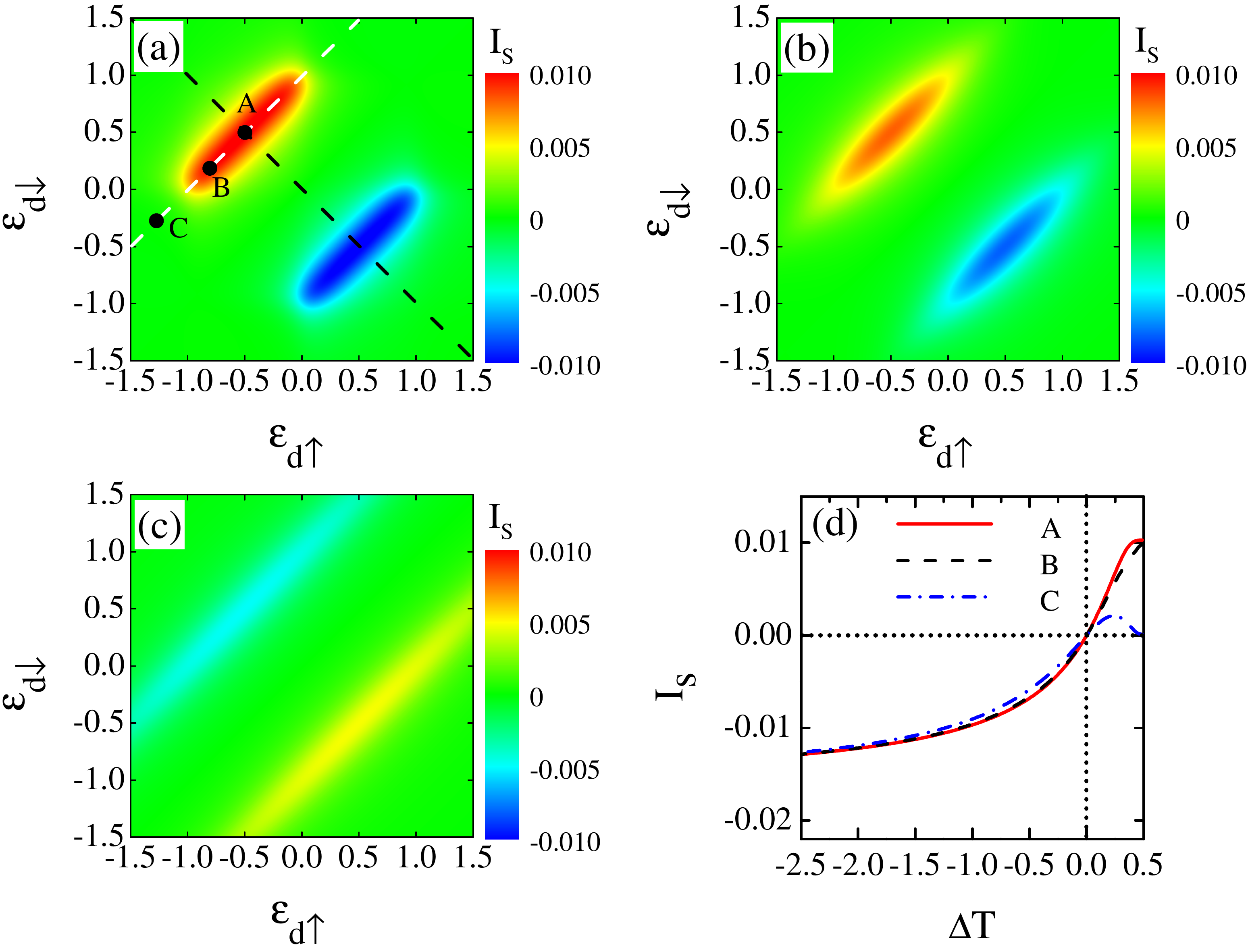}
\centering
\caption{(Color online) Color map of the PSC as a function of $\varepsilon_{d\uparrow}$ and $\varepsilon_{d\downarrow}$ at different electron temperatures as (a) $T_e=0$ ($\Delta T=0.5$), (b) $T_e=0.2$ ($\Delta T=0.3$), and (c) $T_e=0.8$ ($\Delta T=-0.3$). The two dashed lines in (a) indicate the specific sublevel configurations  $\varepsilon_{d\downarrow}=-\varepsilon_{d\uparrow}$ (black dashed line) and $\varepsilon_{d\downarrow}=\varepsilon_{d\uparrow}+\varepsilon_{ph}$ (white dashed line). (d) The evolution of PSC against the temperature difference $\Delta T=T_{b}-T_e$ at three sublevel configurations indicated by the black dots in (a). The parameters used are $\lambda=0.1$, $T_{b}=0.5$, and $\Gamma=0.3$.}\label{fig2}
\end{figure}

\begin{figure}[bp]
\includegraphics[width=\columnwidth]{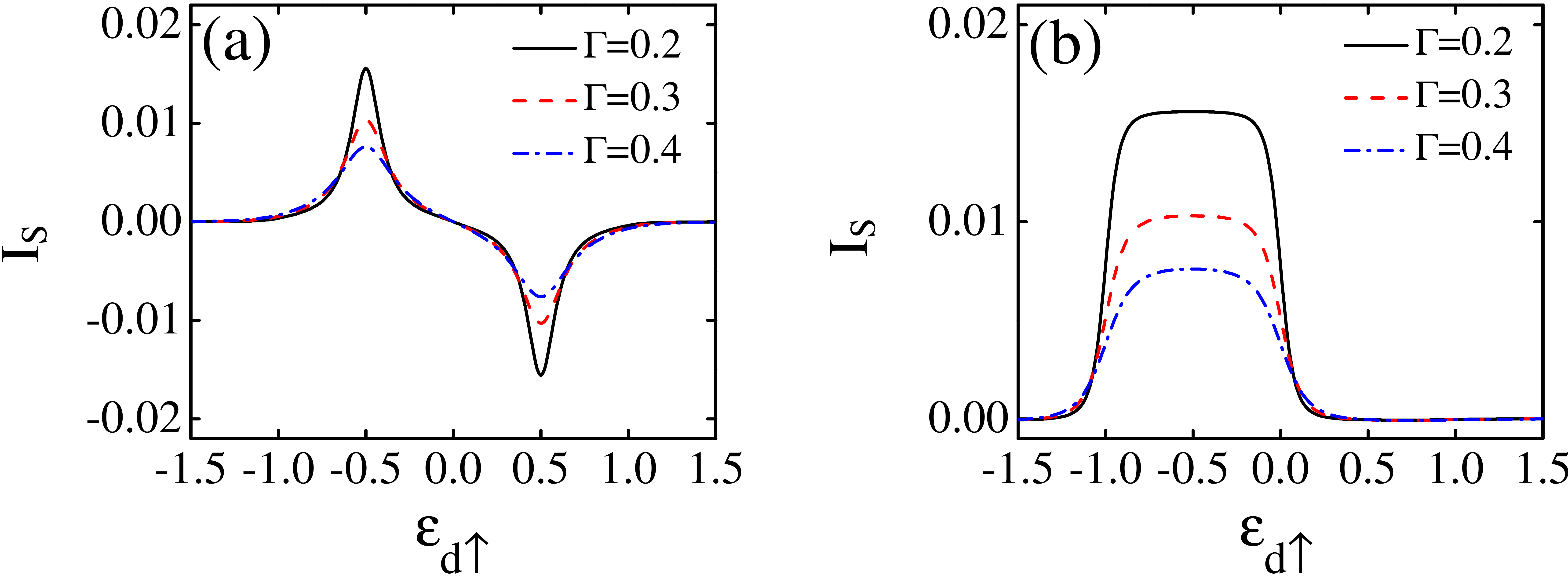}
\centering
\caption{(Color online) Dependence of PSC at $T_e=0$ on the tunneling rate under two specific sublevel configurations as (a) $\varepsilon_{d\downarrow}=-\varepsilon_{d\uparrow}$ and (b) $\varepsilon_{d\downarrow}=\varepsilon_{d\uparrow}+\varepsilon_{ph}$, as indicated by the black and white dashed lines, respectively, in figure \ref{fig2}(a). The parameters used are $\lambda=0.1$ and $T_b=0.5$.}  \label{fig3}
\end{figure}

In this section, we present our numerical results based on equations \,(\ref{eq6}) and (\ref{eq7}). In all the calculations, we take $\varepsilon_{ph}$ as the energy unit. We first display the dependence of PSC on the sublevel configuration at three temperatures as $T_e=0$ (case I), $0<T_e<T_b$ (case II), and $T_b<T_e$ (case III). In case I, two PSC islands like baguettes develop with explicit edges [figure \ref{fig2}(a)]. More precisely, a considerable PSC is generated when the sublevel configuration satisfies the resonance condition
\begin{equation}
|\varepsilon_{d\downarrow}-\varepsilon_{d\uparrow}|=\varepsilon_{ph} \label{rc}
\end{equation}
together with that $\varepsilon_{d\uparrow}$ and $\varepsilon_{d\downarrow}$ have opposite signs. A positive PSC requires that $\varepsilon_{d\uparrow}<0<\varepsilon_{d\downarrow}$ while a negative PSC needs $\varepsilon_{d\downarrow}<0<\varepsilon_{d\uparrow}$, as indicated clearly in figures \ref{fig1}(c) and \ref{fig1}(d). In case II, the visible areas of the PSC islands are enlarged but their edges become obscure [figure \ref{fig2}(b)], which is attributed to the blurred Fermi distribution of the electron in the pole. As for case III, apart from the variations of the current magnitudes and the island areas, the PSCs on the two islands reverse sign simultaneously [figure \ref{fig2}(c)], as a result of the process in figure \ref{fig1}(e) dominating the one in figure \ref{fig1}(c). We show in section \ref{interacting} that, when a strong intradot Coulomb repulsion is considered, there will be two more PSC islands in the color maps, which just correspond to the PSCs established between the higher two sublevels $\varepsilon_{d\uparrow}+U$ and $\varepsilon_{d\downarrow}+U$ when they are adjusted to the vicinity of $\mu_F$, as we discussed in section \ref{intr}. Figures \ref{fig2}(a)-\ref{fig2}(c) indicate that tuning the sublevel configuration along the black dashed line ($\varepsilon_{d\downarrow}=-\varepsilon_{d\uparrow}$) in figure \ref{fig2}(a) is the optimal path for a delicate controlling on the PSC. In figure \ref{fig2}(d), we present the detailed evolution of the PSCs against the temperature difference $\Delta T$ at three sublevel configurations $A$, $B$, and $C$, as indicated by the black dots aligned on the white dashed line ($\varepsilon_{d\downarrow}=\varepsilon_{d\uparrow}+\varepsilon_{ph}$) in figure \ref{fig2}(a). It is observed that the PSCs all collapse to zero at $\Delta T=0$ and converge to a saturation, which are direct results of the third analytical insight mentioned above. Note that the emergence of a common saturation arises from the identical Zeeman splitting of the three selected configurations. Furthermore, it is evident that the dependence of PSCs on the temperature difference is much sensitive at $\Delta T>0$, for the reason that a dramatic change of the Fermi distribution near $\mu_F$ occurs only at low $T_e$. The PSCs at $A$, $B$ show up qualitatively different behaviors from the ones at $C$. This is attributed to the subtle competitions between the two factors, i.e., sublevel configuration in QD and the electron temperature, as depicted in figures \ref{fig1}(c)-\ref{fig1}(e).

\begin{figure}[bp]
\includegraphics[width=\columnwidth]{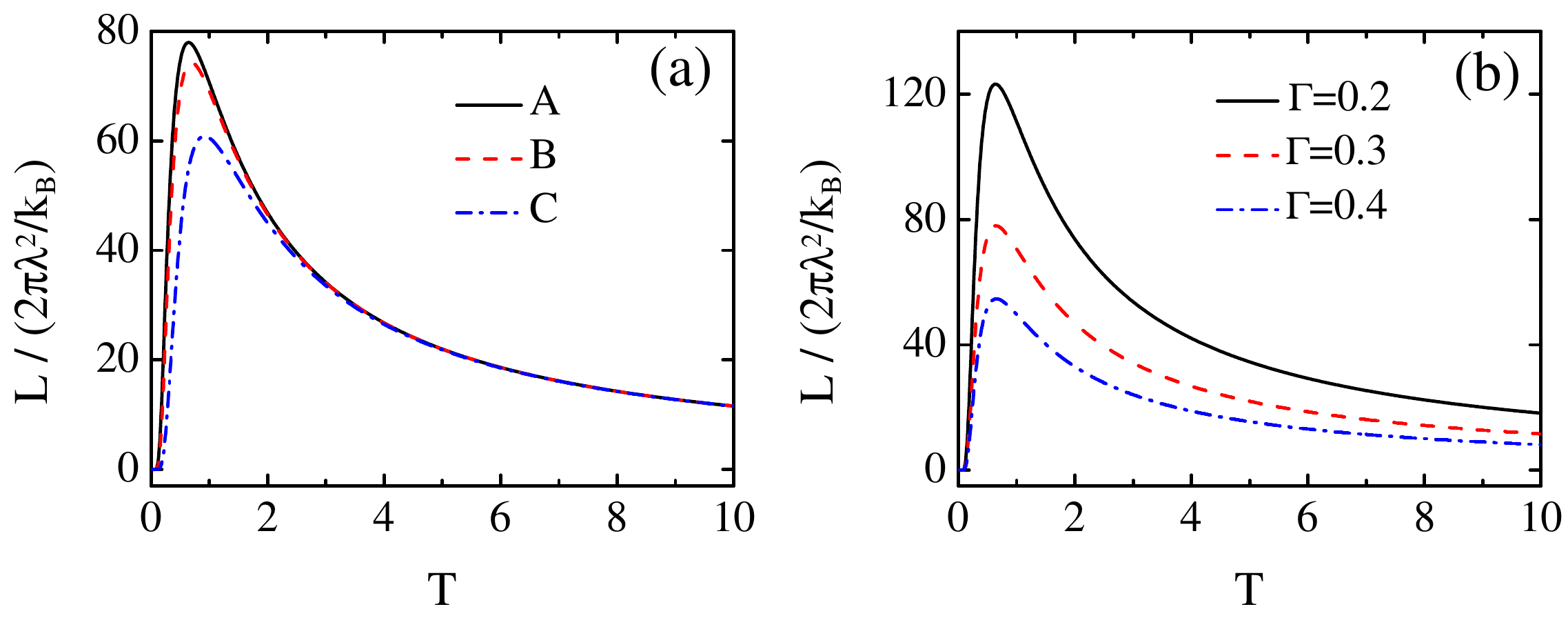}
\centering
\caption{(Color online) (a) The kinetic coefficient $L$ as a function of the temperature $T$ at three sublevel configurations $A$, $B$, and $C$ as indicated in figure \ref{fig2}(a). Here $\Gamma=0.3$. (b) Effect of the tunneling rate $\Gamma$ on the kinetic coefficient $L$ at the sublevel configuration A.}  \label{fig4}
\end{figure}

Now we discuss the dependence of PSC on the tunneling rate. In figure \ref{fig3}(a), the evolution of PSC at $T_e=0$ against the specific sublevel configurations indicated by the black dashed lines in figure \ref{fig2}(a) is traced for various $\Gamma$. As we can see, a strong positive (negative) PSC resonant peak is formed at $\varepsilon_{d\downarrow}-\varepsilon_{d\uparrow}=\varepsilon_{ph}$ ($\varepsilon_{d\uparrow}-\varepsilon_{d\downarrow}=\varepsilon_{ph}$) with its full width at half maximum being $\Gamma$. The broadenings of resonances implicate a tolerance allowed by the resonance condition equation (\ref{rc}) since each of the sublevels is effectively broadened by $\Gamma$. As $\Gamma$ is increased, the absolute maximums of PSC are reduced, which can be traced back to the suppressed local density of states $\rho_\sigma(\omega)$ involved in the function $\mathcal T_\sigma(\omega)$. Physically, the reduction of PSC is due to that the broadenings of sublevels diminish effectively the Zeeman splitting that is important in the resonant spin flip mechanism. On the other hand, for a vanishingly small $\Gamma$ the local density of states $\rho_\sigma(\omega)$ will reduce to the Dirac-$\delta$ function such that the integral in equation (\ref{eq6}) is divergent. This must be incorrect since no electron current and thus no PSC can be set up without a tunneling coupling. We note that the $\Gamma\rightarrow0$ limit should not be taken into account here, since the perturbation theory we performed is on the parameter $\lambda$, which is considered as the smallest energy scale, instead of $\Gamma$. However, thorough insights on the role played by $\Gamma$ is beyond the scope of present work. In figure \ref{fig3}(b), similar dependence on the tunneling rate is shown for the PSCs along another sublevel configurations indicated by the white dashed line in figure \ref{fig2}(a).

We would like to mention the thermoelectric PSC in the linear response regime. To this end, one can expand the current formula equation (\ref{eq6}) to the first order in $\Delta T$ at fixed $T$ ($T\equiv T_e$) and using the relation $I_s=\hbar J_\uparrow$ to obtain $I_s=L \Delta T$, with the kinetic coefficient (see \ref{app})
\begin{equation}
L=2\pi\lambda^{2}\frac{\varepsilon_{ph}N_T}{k_B T^2}\int d\omega\mathcal T_\uparrow(\omega)f(\omega_-)[1-f(\omega_+)],\label{eq12}
\end{equation}
where $N_T$ represents the thermal phonon number $N_{ph}$ at $T_b=T$. In figure \ref{fig4}(a), the dependence of kinetic coefficient on the temperature is plotted at three sublevel configurations $A$, $B$, and $C$ indicated in figure \ref{fig2}(a). One sees that the kinetic coefficient is non-monotonic in the temperature: it increases quickly to the maximum at $T$ about $1$ (i.e., $k_B T\approx\varepsilon_{ph}$), and then decreases more and more slowly as $T$ increases. Particularly, it vanishes at $T=0$, since then no thermal phonons is available. In addition, comparing the maximums of curve $A$, $B$, and $C$, the global optimal kinetic coefficient is predicted to be achieved at the sublevel configurations $\varepsilon_{d\uparrow}=-\varepsilon_{d\downarrow}=\pm\varepsilon_{ph}/2$. In figure \ref{fig4}(b), it is shown that the kinetic coefficient is suppressed as $\Gamma$ increases, which is due to the same reason for the suppression of PSC at large tunneling rate, as mentioned above.

\section{QD with Coulomb interaction}\label{interacting}
Now we turn to validate our discussion in section \ref{intr} that the inclusion of a strong Coulomb interaction in the CNT QD will not disturb the substantial physical scenario, provided that only the Coulomb blockade effect survives at weak tunnel coupling or at high enough temperatures. For this purpose, we would like to calculate the PSC as a function of the sublevels in the presence of a nonzero Coulomb interaction. However, for an interacting QD, the generalized self-consistent Born approximation we employed in section \label{noninteracting} is impracticable since the Wick's theorem can only be applied to a quadratic unperturbed Hamiltonian \cite{Mahan2000}. Therefore, we resort to the simple but useful master equation method \cite{Bruus2004} to incorporate the electron-electron correlations, which is reliable for $\textrm{min}\{k_B T_e, k_B T_b\} \geq \textrm{max}\{\Gamma, \lambda^2\}$.

\subsection{Master equation method}
To proceed, we start with dividing the total Hamiltonian into two parts as $H=H_0+H_{int}$, where $H_0=H_{pole}+H_{ph}+\sum_{\sigma}\varepsilon_{d\sigma}d_{\sigma}^{\dag}d_{\sigma}+Un_{d\uparrow}n_{d\downarrow}$ and
$H_{int}=H_{tunnel}+\lambda(a+a^\dag)\sum_{\sigma}d_\sigma^\dag d_{\bar\sigma}$. In the $H_0$ term, the pole, the vibrational mode, and the QD are independent of each other and thus they are in respective thermal equilibrium states. It is the $H_{int}$ term which couples the QD to the pole and to the vibrational mode that makes the electronic transport between the QD and the pole possible. We assume that both the tunnel coupling and the SVI are so weak that the $H_{int}$ term can be treated as a perturbation within the framework of the master equation method. Here, we will restrict the calculations to lowest nonvanishing order of $H_{int}$, which has been shown to describe the Coulomb blockade quite accurately in an interacting QD connecting two poles but without the SVI \cite{Averin1991,Bonet2002,Muralidharan2006}. The steady-state occupation probabilities $P_m$, $m\in\{0, \uparrow, \downarrow, \uparrow \downarrow\}$, for the QD states are determined by the master equations
\begin{eqnarray}
&&\hspace{-1cm}0=\frac{dP_{0}}{dt}=\gamma _{0\leftarrow \uparrow }P_{\uparrow }+\gamma_{0\leftarrow \downarrow }P_{\downarrow }-( \gamma _{\uparrow\leftarrow 0}+\gamma _{\downarrow \leftarrow 0}) P_{0}, \\
&&\hspace{-1cm}0=\frac{dP_{\uparrow }}{dt}=\gamma _{\uparrow \leftarrow 0}P_{0}+\gamma_{\uparrow \leftarrow \uparrow \downarrow }P_{\uparrow \downarrow}+\gamma _{\uparrow \leftarrow \downarrow }P_{\downarrow }-( \gamma_{0\leftarrow \uparrow }+\gamma _{\uparrow \downarrow \leftarrow \uparrow}+\gamma _{\downarrow \leftarrow \uparrow }) P_{\uparrow }, \\
&&\hspace{-1cm}0=\frac{dP_{\downarrow }}{dt}=\gamma _{\downarrow \leftarrow0}P_{0}+\gamma _{\downarrow \leftarrow \uparrow \downarrow }P_{\uparrow\downarrow }+\gamma _{\downarrow \leftarrow \uparrow }P_{\uparrow}-( \gamma _{0\leftarrow \downarrow }+\gamma_{\uparrow \downarrow\leftarrow \downarrow }+\gamma _{\uparrow \leftarrow \downarrow })P_{\downarrow }, \\
&&\hspace{-1cm}0=\frac{dP_{\uparrow \downarrow }}{dt}=\gamma _{\uparrow \downarrow\leftarrow \uparrow }P_{\uparrow }+\gamma _{\uparrow \downarrow \leftarrow\downarrow }P_{\downarrow }-( \gamma _{\uparrow \leftarrow \uparrow\downarrow }+\gamma _{\downarrow \leftarrow \uparrow \downarrow })P_{\uparrow \downarrow },
\end{eqnarray}
together with the normalization condition $P_{0}+P_{\uparrow }+P_{\downarrow }+P_{\uparrow \downarrow}=1$. The rates for tunneling-induced transition between the states are obtained from the generalized Fermi's golden rule \cite{Bruus2004} as
$\gamma _{\sigma \leftarrow 0}=\hbar^{-1}\Gamma f( \varepsilon _{d\sigma}) $, $\gamma _{\uparrow \downarrow \leftarrow \sigma }=\hbar^{-1}\Gamma f(\varepsilon _{d\bar{\sigma}}+U) $, $\gamma _{0\leftarrow \sigma}=\hbar^{-1}\Gamma [ 1-f( \varepsilon _{d\sigma }) ] $, and $\gamma_{\sigma \leftarrow \uparrow \downarrow }=\hbar^{-1}\Gamma [ 1-f(\varepsilon _{d\bar{\sigma}}+U) ] $. Along the same line, one can derive the SVI-induced spin-flip rates as
\begin{eqnarray}
&&\hspace{-2.4cm}\gamma_{\bar{\sigma}\leftarrow\sigma}=2\pi\hbar^{-1}\sum_{n=0}^{\infty}\rho_{n}\vert \langle n\vert \langle \bar{\sigma}\vert a^{\dag}H_{int}\vert \sigma\rangle \vert n\rangle\vert ^{2}\delta[(  \varepsilon_{d\sigma}+\varepsilon_{ph})  -\varepsilon_{d\bar{\sigma}}]\nonumber\\
&&\hspace{-1cm}+2\pi\hbar^{-1}\sum_{n=0}^{\infty}\rho_{n}\vert \langle n\vert \langle \bar\sigma \vert a H_{int}\vert \sigma\rangle \vert n\rangle \vert ^{2}\delta[  \varepsilon_{d\sigma}-(\varepsilon_{d\bar{\sigma}}+\varepsilon_{ph}) ]\nonumber\\
&&\hspace{-1.4cm}=2\pi\hbar^{-1}\lambda^{2}\sum_{n=0}^{\infty}\rho_{n}[n^{2}\delta(\varepsilon_{d\sigma}+\varepsilon_{ph}  -\varepsilon_{d\bar{\sigma}})+(1+n)^{2}\delta( \varepsilon_{d\sigma}-\varepsilon_{d\bar{\sigma}}-\varepsilon_{ph} )]  \nonumber\\
&&\hspace{-1.4cm}=2\pi\hbar^{-1}\lambda^{2}(1+2N_{ph})[N_{ph}\delta(\varepsilon_{d\sigma}-\varepsilon_{d\bar{\sigma}}+\varepsilon_{ph})+(N_{ph}+1)\delta(\varepsilon_{d\sigma}-\varepsilon_{d\bar{\sigma}}-\varepsilon_{ph}) ],
\end{eqnarray}
where $\vert n\rangle$ denotes the Fock state occupied by $n$ phonons and $\rho_{n}=Z^{-1}e^{-n\varepsilon_{ph}/k_B T_b}$ is its weight factor with $Z=\sum_{n=0}^{\infty}e^{-n\varepsilon_{ph}/k_B T_b}$ being the partition function. For the convenient of practical calculations, we would like to replace the $\delta(x)$ function by a Lorentztian function $\frac{\eta/\pi}{x^2+\eta^2}$ with a small width $\eta$. From the solution to the master equations, the spin-dependent electron current flowing through pole into the QD is obtained as
\begin{equation}
J_{\sigma }=\gamma _{\sigma \leftarrow 0}P_{0}+\gamma_{\uparrow \downarrow \leftarrow \bar{\sigma}}P_{\bar{\sigma}}-\gamma_{0\leftarrow \sigma }P_{\sigma }-\gamma _{\bar{\sigma}\leftarrow \uparrow \downarrow }P_{\uparrow \downarrow}.\label{recurrent}
\end{equation}
By the numerical calculations, we confirm that the charge current conservation law $J_\uparrow+J_\downarrow=0$ (therefore, $I_s=\hbar J_\uparrow$) and that the charge current vanishes at $T_e=T_b$ are both respected within the master equation method.

\begin{figure}[bp]
\includegraphics[width=\columnwidth]{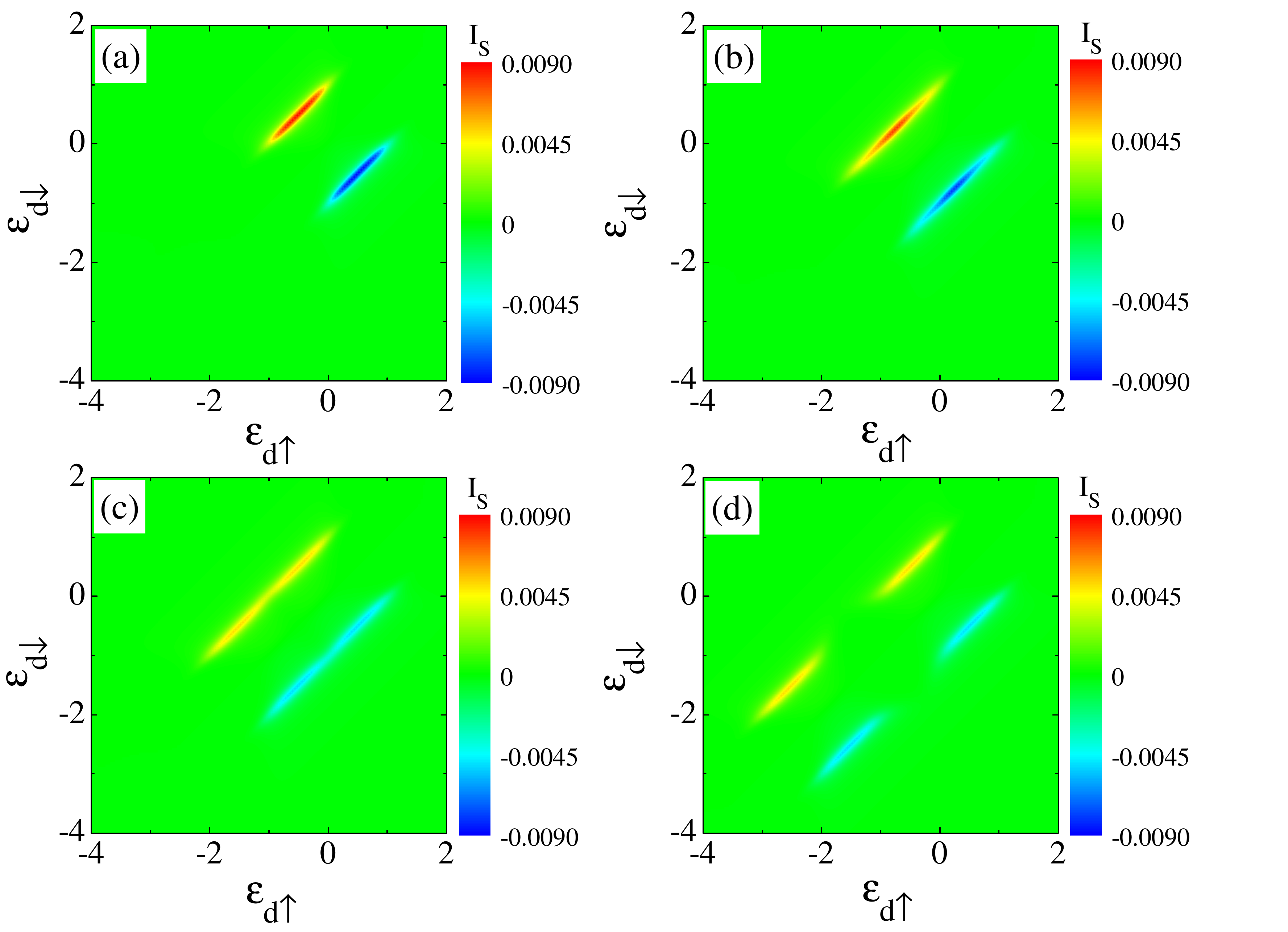}
\centering
\caption{(Color online) Color map of the PSC as a function of $\varepsilon_{d\uparrow}$ and $\varepsilon_{d\downarrow}$ in the case $T_e<T_b$ ($T_e=0.1$, $T_b=0.5$) with different Coulomb interaction (a) $U=0$, (b) $U=0.5$, (c) $U=1$, and (d) $U=2$. Other parameters are $\lambda=0.1$, $\Gamma=0.05$, and $\eta=0.02$.}  \label{figapp1}
\end{figure}

\subsection{Numerical results and discussions}
In figure \ref{figapp1}, we show how the Coulomb interaction affect the dependence of PSCs on the sublevels in the case $T_e<T_b$. In the absence of the Coulomb interaction [figure \ref{figapp1}(a)], there exists two PSC islands which is consistent with figures \ref{fig2}(a) and \ref{fig2}(b). For a weak but nonzero $U$ [figure \ref{figapp1}(b)], the PSC islands are stretched and the absolute PSCs are suppressed at the same time. Once $U$ becomes larger than the phonon energy [figures \ref{figapp1}(c) and \ref{figapp1}(d)], each PSC island is divided into two isolate islands. The emergent islands just correspond to the PSCs established between the two sublevels $\varepsilon_{d\uparrow}+U$ and $\varepsilon_{d\downarrow}+U$ when they are adjusted to the vicinity of $\mu_F$, as we discussed in section \ref{intr}. We further explain this scenario in detail, for example, under the sublevel configuration $\varepsilon_{d\uparrow}<\varepsilon_{d\downarrow}<\varepsilon_{d\uparrow}+U<\mu_F<\varepsilon_{d\downarrow}+U$. When the sublevels are off resonance (i.e., $\varepsilon_{d\downarrow}-\varepsilon_{d\uparrow}\ne\varepsilon_{ph}$), the QD is almost occupied by one spin-up electron and the Coulomb repulsion prohibits the pole electrons from entering the QD. Therefore no PSC flows in the pole. On the other hand, when the sublevels are on resonance (i.e., $\varepsilon_{d\downarrow}-\varepsilon_{d\uparrow}=\varepsilon_{ph}$), the trapped spin-up QD electron has the possibility to reverse its spin and transit to the spin-down sublevel by absorbing one phonon, which then allows a spin-up pole electron with energy $\varepsilon_{d\uparrow}+U$ to enter the QD. Subsequently, the injected QD electron can absorb one phonon and transit to the sublevel $\varepsilon_{d\downarrow}+U$. Finally, it tunnels easily to the pole since $\varepsilon_{d\downarrow}+U>\mu_F$. Theses combined tunneling processes are responsible for the emergent positive PSC island. One can also notice that the absolute maximum of the PSC in figure \ref{figapp1}(d) is roughly reduced by half in comparison to the one in figure \ref{figapp1}(a), which could be explained as follows. We first focus on the two PSC islands at the top right corner in figure \ref{figapp1}(d). As illustrated in figure \ref{fig1}(c) for $U=0$, when a spin-up electron at the lower sublevel transits to the upper sublevel by absorbing a phonon, another spin-up pole electron can tunnel into the lower sublevel immediately. However, for a strong Coulomb repulsion, the latter process is blockaded due to the presence of the spin-down QD electron at the upper sublevel. Only if that QD electron tunneling out to the pole will allow a spin-up pole electron to enter the QD. Mathematically, the reduction of PSC is attributed to the fact that the second term in equation (\ref{recurrent}) which contributes equal as the first term to the total current in the $U=0$ case vanishes for a strong $U$, while the last two terms in equation (\ref{recurrent}) are always vanishingly small for the parameters used in figure \ref{figapp1}. As for the two emergent PSC islands, the relatively smaller current amplitude is due to the small empty state occupation probability $P_0$.

\begin{figure}[bp]
\includegraphics[width=\columnwidth]{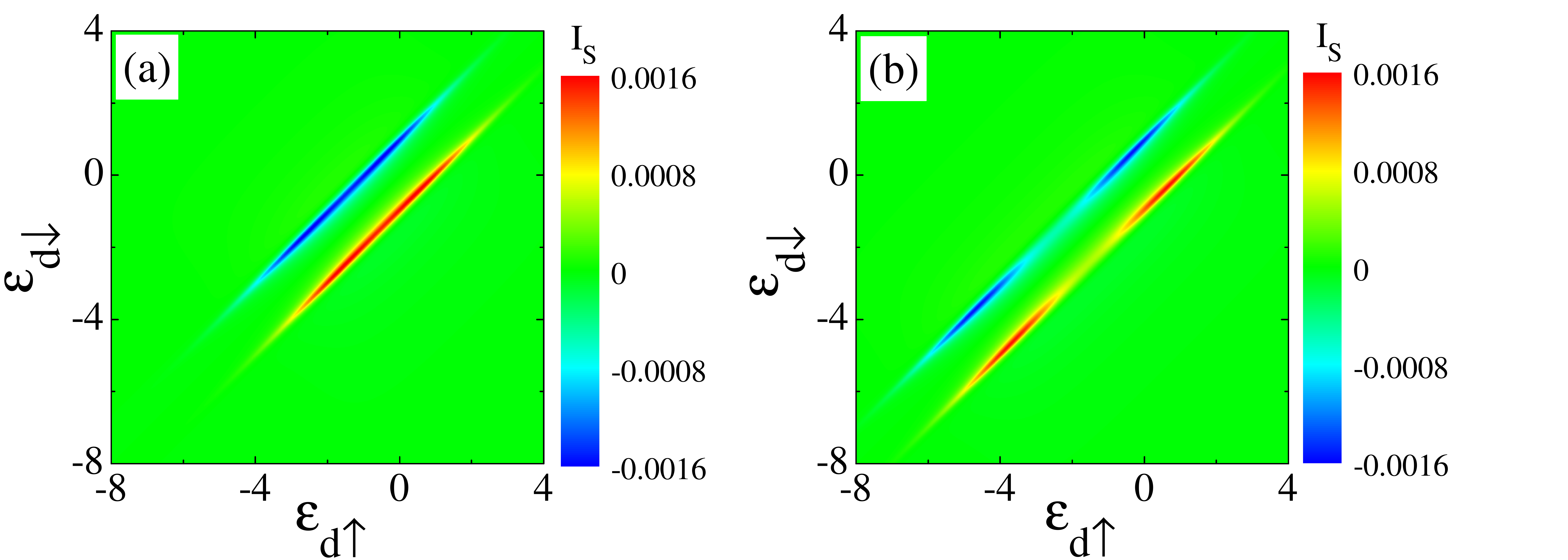}
\centering
\caption{(Color online) Color map of the PSC as a function of $\varepsilon_{d\uparrow}$ and $\varepsilon_{d\downarrow}$ in the case $T_e>T_b$ ($T_e=0.7$, $T_b=0.5$) with different Coulomb interaction (a) $U=2$ and (b) $U=4$. Other parameters are $\lambda=0.1$, $\Gamma=0.05$, and $\eta=0.02$.}  \label{figapp2}
\end{figure}

Figure \ref{figapp2} shows the dependence of PSCs on the sublevel configuration in the case $T_e>T_b$ with finite $U$. In contrast to figure \ref{figapp1}, the PSCs reverse direction as the electron temperature varies from $T_e<T_b$ to $T_e>T_b$, which is in agreement with the $U=0$ case as shown in figure \ref{fig2}. In figure \ref{figapp2}(a), the two emergent PSC islands induced by the Coulomb interaction are merged with the original islands due to the smeared Fermi distribution around $\mu_F$ at high $T_e$. However, they manifest themselves at a larger $U$, as shown in figure \ref{figapp2}(b).

\section{Conclusions}\label{summary}
In conclusion, we have found that a QD formed in a suspended CNT exposed to an external magnetic field could act as a thermoelectric unipolar SB which generates PSC. In this setup, the spin flip source is natural due to the interplay between the intrinsic spin-orbit coupling and the vibrational modes of the suspended CNT, rather than the previous ones induced by the somewhat complicated time-varying external fields \cite{Brataas2002,Zhang2003,Wang2003}. Moreover, the driving force of this SB setup is a consequence of the nontrivial thermoelectric effect with respect to the Fermi (pole) and Bose (thermal bath) reservoirs. The magnitude and the direction of the generated PSC are dependent on such four factors as i) the strength of SVI, ii) the sublevel configuration in QD, iii) the electron ($T_e$) and bath ($T_{b}$) temperatures, and iv) the tunnelling rate between the QD and the pole. In particular, for finite temperature difference between the pole and the thermal bath, a joint adjustment on the sublevel configuration and the tunneling rate suffices the delicate controlling on the PSC. On the experimental aspect, the sublevels in a CNT QD is finely tunable nowadays by the interplay of a gate voltage and an external applied magnetic field \cite{Jarillo2005,Kuemmeth2008,Fang2008,Churchill2009}, and the tunneling rate can also be conveniently regulated by a gated tunneling barrier. In addition, in the linear response regime, it is found that the kinetic coefficient is non-monotonic in the temperature $T$ and it reaches its maximum when $k_B T$ is about one phonon energy. We have also demonstrated that the existence of a strong intradot Coulomb interaction is irrelevant for our SB, provided that high-order cotunneling processes are suppressed. Obviously, the SB setup we addressed in this work explicitly indicates a potential application of the versatile CNTs. We hope that our results could be helpful for obtaining controllable PSC, which plays a significant role in spintronics.

\section{Acknowledgments}
This work is supported by NSFC (Grants No.\,11325417 and No.\,11674139) and PCSIRT (Grant No.\,IRT1251).

\appendix
\section{Derivation of equation (\ref{eq12})}\label{app}
In the spin-dependent current formula equation (\ref{eq6}), while $\mathcal T_\sigma(\omega)$ is independent of the temperatures $\mathcal{F}(\omega)$ is actually a function of $T_e$ and $T_b$ as
\begin{equation}
\hspace{-2.3cm}\mathcal F(\omega,T_{b},T_e)=[1+N_{ph}(T_{b})] [f(\omega_{-},T_e)-1]  f(\omega_{+},T_e)-N_{ph}(T_{b})f(\omega_{-},T_e)[f(\omega_{+},T_e)-1].\label{eqA1}
\end{equation}
In the linear response regime, we keep the Taylor expansion of $\mathcal{F}(\omega)$ to the first order in the temperature difference $\Delta T=T_b-T$, with $T\equiv T_e$ denotes the fixed electron temperature, as
\begin{equation}
\mathcal F(\omega,T_{b},T)=\mathcal F(\omega,T_{b},T)\mid_{T_{b}=T}+\frac{\partial \mathcal F(\omega,T_{b},T)}{\partial T_{b}}\mid_{T_{b}=T}\Delta T.\label{eqA2}
\end{equation}
Notice the identity
\begin{equation}
[1+N_{ph}(T)] [f(\omega_{-},T)-1]f(\omega_{+},T)=N_{ph}(T)f(\omega_{-},T)[f(\omega_{+},T)-1],\label{eqA3}
\end{equation}
the first term in equation (\ref{eqA2}) vanishes exactly. The second term in equation (\ref{eqA2}) reads
\begin{eqnarray}
&&\frac{\partial F(\omega,T_{b},T)}{\partial T_{b}}\mid_{T_{b}=T}\Delta T\nonumber\\
&&\hspace{-0.5cm}=\frac{dN_{ph}(T_{b})}{dT_{b}}\mid_{T_{b}=T}\{[f(\omega_{-},T)-1]f(\omega_{+},T)-f(\omega_{-},T)[f(\omega_{+},T)-1]\}\Delta T\nonumber\\
&&\hspace{-0.5cm}=\frac{\varepsilon_{ph}N_{ph}(T)}{k_{B}T^{2}}\{[1+N_{ph}(T)][f(\omega_{-},T)-1]f(\omega_{+},T)\nonumber\\
&&\hspace{2.5cm}-[1+N_{ph}(T)]f(\omega_{-},T)[f(\omega_{+},T)-1]\}\Delta T,\label{eqA4}
\end{eqnarray}
Applying equation (\ref{eqA3}) to the first term in the brace of equation (\ref{eqA4}) one immediately obtains
\begin{equation}
\frac{\partial F(\omega,T_{b},T)}{\partial T_{b}}\mid_{T_{b}=T}\Delta T=\frac{\varepsilon_{ph}N_{ph}(T)}{k_{B}T^{2}}f(\omega_{-},T)[1-f(\omega_{+},T)]\Delta T.\label{eqA5}
\end{equation}
Collecting equations (\ref{eqA2}) and (\ref{eqA5}) into equation (\ref{eq6}) and using the relation $I_s=\hbar J_\uparrow$ one finally obtains the equation (\ref{eq12}) in the main text.

\end{document}